\documentclass[prd,twocolumn,amsmath,amssymb]{revtex4}
\usepackage{graphicx}

\begin{document}

\title{\boldmath Measurements of the cross sections for $e^+e^- \rightarrow
{\rm hadrons}$ at 3.650, 3.6648, 3.773 GeV and the branching
fraction for
$\psi(3770)\rightarrow {\rm non-}D\bar D$}
\author{
\begin{small}
M.~Ablikim$^{1}$,      J.~Z.~Bai$^{1}$,            Y.~Ban$^{11}$,
J.~G.~Bian$^{1}$,      X.~Cai$^{1}$,               H.~F.~Chen$^{15}$,
H.~S.~Chen$^{1}$,      H.~X.~Chen$^{1}$,           J.~C.~Chen$^{1}$, 
Jin~Chen$^{1}$
% ...
\end{small}
\\(BES Collaboration)\\
}
\vspace{0.2cm}
\affiliation{ 
\begin{minipage}{145mm}
$^{1}$ Institute of High Energy Physics, Beijing 100049, People's Republic
of China\\
% ...
$^{11}$ Peking University, Beijing 100871, People's Republic of China\\
% ...
$^{15}$ University of Science and Technology of China, Hefei 230026,
People's Republic of China\\
% ...
\vspace{0.4cm}
\end{minipage}
}

\begin{abstract}

   Using the BES-II detector at the BEPC Collider, we measured 
the lowest order cross sections and the $R$ values
($R=\sigma^0_{e^+e^- \rightarrow {\rm hadrons}}/
\sigma^0_{e^+e^- \rightarrow \mu^+\mu^-}$) 
for inclusive hadronic event production
at the center-of-mass energies
of 3.650 GeV, 3.6648 GeV and 3.773 GeV. 
The results lead to $\bar R_{uds}=2.224\pm 0.019\pm 0.089$
which is the average of these measured 
at 3.650 GeV and 3.6648 GeV,
and $R=3.793\pm 0.037 \pm 0.190$
at $\sqrt{s}=3.773$ GeV.
We determined the lowest order cross section for $\psi(3770)$ production to be 
$\sigma^{\rm B}_{\psi(3770)} = (9.575\pm 0.256 \pm 0.813)~{\rm nb}$ at 3.773 GeV,
the branching fractions for $\psi(3770)$ decays to be
$BF(\psi(3770) \rightarrow D^0\bar D^0)=(48.9 \pm 1.2 \pm 3.8)\%$,
$BF(\psi(3770) \rightarrow D^+ D^-)=(35.0 \pm 1.1 \pm 3.3)\%$ and
$BF(\psi(3770) \rightarrow D\bar{D})=(83.9 \pm 1.6 \pm 5.7)\%$,
which result in the total non-$D\bar D$ branching fraction of $\psi(3770)$ decay to be
$BF(\psi(3770) \rightarrow {\rm non}-D\bar D)=(16.1 \pm 1.6 \pm 5.7)\%$.
\end{abstract}

\maketitle

\section{Introduction}
The established picture of hadron production
by $e^+e^-$ annihilation in continuum region is that 
the annihilation proceeds via quark-antiquark pair
production  where the photon couples directly to the charge
of the pointlike quarks.
A consequence of this picture is that the total
lowest order cross section, $\sigma^{\rm B}_{had}$, 
for inclusive hadron production in $e^+e^-$ annihilation
must be proportional to the lowest order cross section, 
$\sigma^{\rm B}_{\mu^+\mu^-}$,
for muon pair production, which results in the relation
\begin{equation}
{\sigma^{\rm B}_{\rm had}}= 
   3 \sum_i^{N_f} Q_i^2~\sigma^{\rm B}_{\mu^+\mu^-},
\end{equation}
where $Q_i$ is the charge of the 
i$th$ quark;
the factor of 3
accounts for three different colors of quarks;
the sum runs over all quark flavors, $N_f$, 
involved, for which the quark pair production thresholds are below the 
$e^+e^-$ annihilation energy.
The Eq. (1) indicates the ratio
\begin{equation}
R=\frac{\sigma^{\rm B}_{\rm had}}
        {\sigma^{\rm B}_{\mu^+\mu^-}}=3 \sum_i^{N_f} Q_i^2
\end{equation}
to be constant as long as the c.m. (center-of-mass) energy $E_{cm}$ 
does not overlap with resonances
or the threshold of the production of new quark flavors.
It also indicates that the $R_{uds}$ value 
for continuum light hadron (containing u, d and s quarks) production
should tend to be constant in the energy region above 2 GeV.

This naive theoretical prediction for the $R$ value has to be modified to
take into account the finite mass of the quarks and the emission
of the gluons by the quarks. 
In principle, the $R$ values can be computed in the pQCD (perturbative QCD) 
with these corrections.
So precise measurements of $R$ values at low energy region
are important for the test of 
the prediction by the pQCD in this energy region.
Moreover the $R$ values at all energies are needed
to calculate the effects of vacuum polarization on the parameters
of the Standard Model. 
For example,
the dominant uncertainties in the quantities $\alpha(M^2_Z)$,
the QED running coupling constant 
evaluated at the mass of $Z^0$,
and $a_{\mu}=(g-2)/2$, the anomalous magnetic moment of the muon,
are due to the calculation of hadronic vacuum polarization\cite{davier}.
A large part of uncertainty in the calculation 
arises from the uncertainties in the measured inclusive hadronic
cross sections in open charm threshold region, in which many resonances
overlap. To get credible measurements of $R$ 
and various lowest order cross sections in this region,
the overlapping effects have to be clarified clearly.

On the other hand, the measurements of the $R$ values 
below and above the threshold of $D\bar D$ production
can be used to determine the branching fractions for 
$\psi(3770) \rightarrow D^0\bar D^0, D^+D^-, D\bar D$, 
and for  $\psi(3770) \rightarrow {\rm non-}D\bar D$
with the measured cross sections for the $D^0\bar D^0$ and
$D^+D^-$ together.
The $\psi(3770)$ resonance is believed to decay predominantly into
$D\bar D$~\cite{Bacino}. 
However, there are discrepancies between the measurements of the $D\bar D$
cross section and the measurements of $\psi(3770)$ cross section which
can be obtained from $\psi(3770)$ resonance parameters. 
In recent days, there are some
publications to report the observation of ${\rm non-}D\bar D$ decays 
of $\psi(3770)$ 
resonance~\cite{bes_non_ddbar1}\cite{bes_non_ddbar2}\cite{cleo}.
In this Letter, we present 
more precise measurements of the $R$ values
at the c.m. energies of 3.650, 3.6648 and 3.773 GeV. 
With the measured $R$ values and the previously measured cross sections
for $D\bar D$ production at 3.773 GeV~\cite{xsct_ddbar_bes},
we determine the branching fractions
for $\psi(3770) \rightarrow D^0\bar D^0, D^+D^-, D\bar D$,
and for $\psi(3770) \rightarrow {\rm non-} D\bar D$

\section{BES-II Detector}

The BES-II is a conventional cylindrical magnetic detector that is
described in detail in Ref.~\cite{BES-II}.  A 12-layer Vertex Chamber
(VC) surrounding the beryllium beam pipe provides input to the event
trigger, as well as coordinate information.  A forty-layer main drift
chamber (MDC) located just outside the VC yields precise measurements
of charged particle trajectories with a solid angle coverage of $85\%$
of $4\pi$; it also provides ionization energy loss ($dE/dx$)
measurements which are used for particle identification.  Momentum
resolution of $1.7\%\sqrt{1+p^2}$ ($p$ in GeV/c) and $dE/dx$
resolution of $8.5\%$ for Bhabha scattering electrons are obtained for
the data taken at $\sqrt{s}=3.773$ GeV. An array of 48 scintillation
counters surrounding the MDC measures the time of flight (TOF) of
charged particles with a resolution of about 180 ps for electrons.
Outside the TOF, a 12 radiation length, lead-gas barrel shower counter
(BSC), operating in limited streamer mode, measures the energies of
electrons and photons over $80\%$ of the total solid angle with an
energy resolution of $\sigma_E/E=0.22/\sqrt{E}$ ($E$ in GeV) and spatial
resolutions of
$\sigma_{\phi}=7.9$ mrad and $\sigma_Z=2.3$ cm for
electrons. A solenoidal magnet outside the BSC provides a 0.4 T
magnetic field in the central tracking region of the detector. Three
double-layer muon counters instrument the magnet flux return and serve
to identify muons with momentum greater than 500 MeV/c. They cover
$68\%$ of the total solid angle.

\section{Measurement of the observed hadronic cross sections}

For a sample of data taken at c.m. energy $E_{{\rm cm},i}$,
the observed cross section for the inclusive hadronic event production
is obtained by
\begin{equation}
\sigma^{\rm obs}(E_{{\rm cm},i}) =\frac{N_{\rm had}(E_{{\rm cm},i})}
                {L(E_{{\rm cm},i})~ \epsilon_{\rm had}(E_{{\rm cm},i})
                ~\epsilon_{\rm had}^{\rm trig}}, 
\end{equation}
where $i$ is the i$th$
energy point at which the data were collected,
${N_{\rm had}(E_{{\rm cm},i})}$ is 
the number of the inclusive hadronic events
observed at this energy; 
$L(E_{{\rm cm},i})$ is the integrated luminosity 
of the data collected;
$\epsilon_{\rm had}(E_{{\rm cm},i})$ is the efficiency for detection
of the inclusive hadronic events, and
$\epsilon_{\rm had}^{\rm trig}$ is the trigger efficiency
for collecting the hadronic events in on-line data acquisition.

\subsection{Measurement of luminosity}
The integrated luminosities of the data sets are determined by
\begin{equation}
L(E_{{\rm cm},i}) =\frac{N_{e^+e^-}(E_{{\rm cm},i}) - n_{b} }
                {\sigma_{e^+e^-}(E_{{\rm cm},i})~ 
                \epsilon_{e^+e^-}(E_{{\rm cm},i})
                ~\epsilon_{e^+e^-}^{\rm trig}},
\end{equation}
where $N_{e^+e^-}(E_{cm,i})$ and $n_{b}$ are the number of the selected 
Bhabha events and
the number of background events respectively,
$\epsilon_{e^+e^-}(E_{cm,i})$ 
is the efficiency for detection of the 
Bhabha events, 
$\epsilon^{\rm trig}_{e^+e^-}$ is the trigger efficiency
for collecting the Bhabha events in on-line data acquisition.
For the data used in the analysis, the trigger efficiency is 
$\epsilon_{e^+e^-}^{\rm trig}=(100.0^{+0.0}_{-0.5})\%$
(see subsection D).

\begin{figure}
\includegraphics[width=9.5cm,height=8.0cm]
{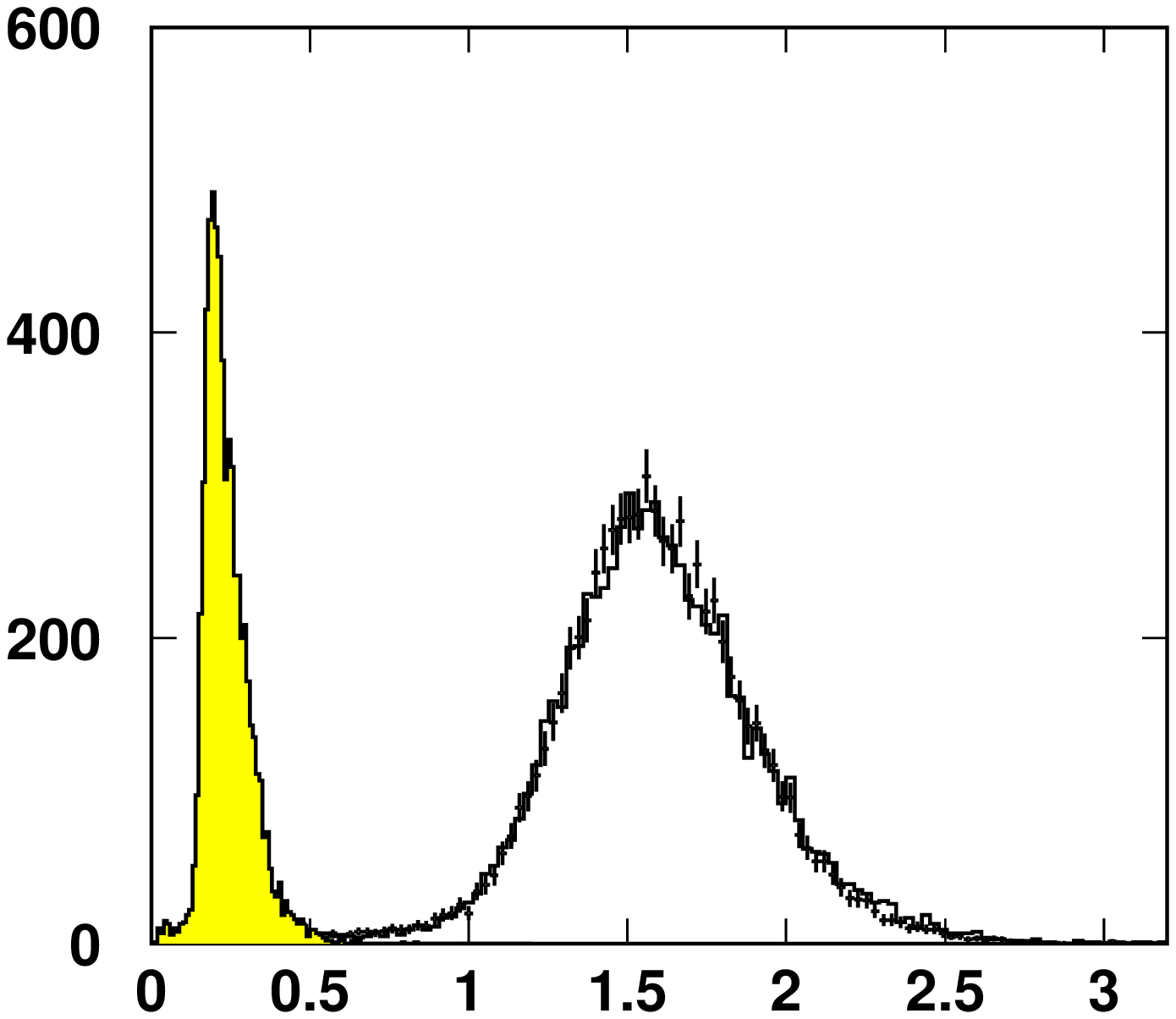}
\put(-160.0,0.0){\Large{$E$~~~~[GeV]}}
\put(-280,65){\rotatebox{90}{\bf\large Number of Events}}
\caption{The distribution of energies deposited for muons (hatched histogram) and
electrons or positrons (points with error bars) in BSC, 
where the sample of the muons and the
electrons or positrons is from the decays of
$\psi(2S) \rightarrow J/\psi \pi^+\pi^-$,
and $J/\psi \rightarrow \mu^+\mu^-~{\rm or}~e^+e^-$; the open histogram is
for the electrons or positrons from the Monte Carlo events of
$\psi(2S) \rightarrow J/\psi \pi^+\pi^-$, and $J/\psi \rightarrow e^+e^-$.}
\label{ebsc}
\end{figure}

To select the candidates for Bhabha scattering 
$e^+e^- \rightarrow (\gamma) e^+e^-$, it is first
required that exactly two
charged tracks with total charge zero be well reconstructed. For each track,
the point of the closest approach to the beam line  must have
the ${\rm radius} \le 1.5~{\rm cm}$
and $|z| \le 15~{\rm cm}$ where $|z|$ is measured
along the beam line from the nominal beam crossing point.
Furthermore, each track is required to
satisfy $|\cos\theta| \le 0.7$, where $\theta$ is the polar angle
of the charged track, to ensure that it is contained within the barrel
region of the detector. Next, it is required that the energy deposited for
each charged track in BSC 
be greater than 1.1 GeV (i.e. $E_{\rm BSC}^{\rm track}>1.1$ GeV)
and at least the magnitude of one charged
track momentum be greater than  $0.9E_{\rm b}$, 
where $E_{\rm b}$ is the beam energy.
Figure~\ref{ebsc} shows the 
distribution of the energies deposited for muons (hatched histogram) and
electrons or positrons (points with error bars) in the BSC, 
where the data sample of the muons and the
electrons or positrons are selected from the decays of
$\psi(2S) \rightarrow J/\psi \pi^+\pi^-$,
and $J/\psi \rightarrow \mu^+\mu^-~{\rm or}~e^+e^-$.
From Fig.~\ref{ebsc} one can see 
that the criterion $E_{\rm BSC}^{\rm track}>1.1$ GeV separates the
$e^+e^- \rightarrow (\gamma) \mu^+\mu^-$ from the Bhabha scattering
effectively. 
In addition, because the Monte Carlo simulation does not
model the energy deposited well in the rib regions of the BSC, 
any charged track from the selected Bhabha events is required 
to hit one of the four regions of the BSC (selected $z$ regions in BSC):
~(1)~$z_{\rm BSC} \le -1.04$ m,
~(2)~$-0.77~{\rm m} \le z_{\rm BSC} \le -0.1$ m,
~(3)~$0.1~{\rm  m} \le z_{\rm BSC} \le 0.77$ m,
~(4)~$z_{\rm BSC} \ge 1.04$ m.

\begin{figure}
\includegraphics[width=9.5cm,height=8.0cm]
{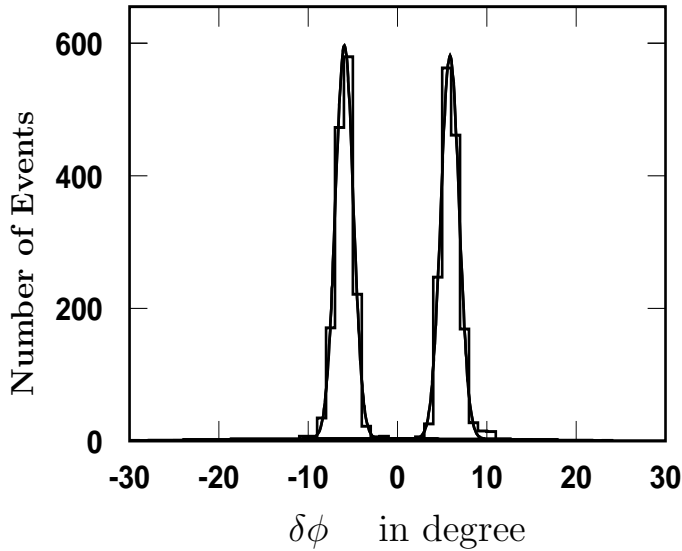}
\put(-165.0,0.0){\Large{$\delta\phi$~~~~in~degree}}
\put(-270,55){\rotatebox{90}{\bf\large Number of Events}}
\caption{The distribution of the $\delta\phi$
($\delta\phi=|\phi_1-\phi_2|-180^o$) of the
selected $e^+$ and $e^-$ tracks.}
\label{bbdltphi}
\end{figure}

    The two oppositely charged tracks go in opposite directions 
in the $R-\phi$ plane.
Because the tracks are bent in the magnetic field, the positions of the two shower
clusters
in the $R-\phi$ plane of the BSC are deviated from the back-to-back directions.
We define the angle difference of the two clusters by
$\delta\phi=|\phi_1-\phi_2|-180^o$ in degrees, where the $\phi_1$ and
$\phi_2$ are the azimuthal angles of the two clusters in the BSC. Figure~\ref{bbdltphi} 
shows the $\delta\phi$ distribution for the events
which satisfy the selection criteria for the Bhabha events.
These events are from a portion of the data taken at 3.773 GeV.
Using a double Gaussian function 
plus a second order polynomial to fit the distribution, 
we obtain  the number of the candidates for $e^+e^- \rightarrow (\gamma) e^+e^-$.
The accepted candidate events are examined 
for background contaminations by visual scan.
The detailed scans for the accepted
$e^+e^- \rightarrow (\gamma) e^+e^-$ events
show that about $0.5\%$ of the accepted events are due to background
contamination. After subtracting the background, 
the pure number of $e^+e^- \rightarrow (\gamma) e^+e^-$ events is retained.

The detection efficiency $\epsilon_{e^+e^-}$ for the Bhabha scattering
$e^+e^- \rightarrow (\gamma) e^+e^-$ 
is determined by analyzing the Monte Carlo events of 
$e^+e^- \rightarrow (\gamma) e^+e^-$. These events are generated
with the 
radiative Bhabha generator~\cite{kleiss_radee} 
written by R. Kleiss {\it et al.}, 
which includes hard photon emission and $\alpha^3$
radiative correction.

Using the pure number selected, 
the visible cross section $\sigma_{e^+e^-}$ read from the generator,
the detection efficiency for $e^+e^- \rightarrow (\gamma) e^+e^-$
obtained by Monte Carlo simulation, and the trigger efficiency,
we can determine the integrated luminosity of the data from Eq. (4).
Applying the procedure to the data sets taken at the three
energy points, we get the measured 
integrated luminosities 
of the data sets.
The second column of table~\ref{lum_evt_list} lists the
integrated luminosities of the data sets, where the errors
are combined from statistical and systematic errors.

The systematic uncertainty in the measured values of the luminosities
arises mainly from the difference between the data and Monte Carlo
simulation. 
Table~\ref{uncertainty_ee} summarizes the systematic uncertainties
due to the $e^+e^-\rightarrow (\gamma)e^+e^-$ event selection criteria. 
The total uncertainty in the measured luminosity is estimated to be
about $1.8\%$.

\begin{table}
\centering
\caption{Summary of the luminosities of the data sets,
the numbers of the selected candidates for
$e^+e^- \rightarrow {hadrons}$ and
the estimated numbers of the events
of the processes
$e^+e^- \rightarrow l^+l^-$ ($l=\tau,e,\mu$),
$e^+e^- \rightarrow e^+e^-l^+l^-$ and
$e^+e^- \rightarrow e^+e^-~hadrons$ 
which were misidentified as the events of
$e^+e^- \rightarrow {hadrons}$. 
}
\label{lum_evt_list}
\begin{tabular}{ccccc} \hline \hline
$E_{cm}$ & L & $N_{\rm had}^{\rm zfit}$ &
$n_{l^+l^-}$ & $n_{e^+e^-l^+l^-}$ \\
(GeV)  &  [${\rm nb^{-1}}$]  &   &   &  $\&$~$n_{e^+e^-h}$  \\ \hline
 3.650 &  $5537.7 \pm 102.3$  & $54576 \pm 239$ &  2038 & 219
\\
 3.6648 & $998.2 \pm $ 19.2 & $9615 \pm 100$ &  382  &  40
\\
 3.773 & $17300.0 \pm 319.6$ & $274021 \pm 538$ &  8603 & 701
\\
\hline \hline
\end{tabular}
\end{table}

\begin{table}
\centering
\vspace{0.2cm}
\caption{The relative systematic uncertainties in the measured luminosity
due to the $e^+e^-\rightarrow (\gamma)e^+e^-$ event selection criteria.}
\label{uncertainty_ee}
\begin{tabular}{lr}
 \hline \hline
Criterion \hspace{5.5cm}          & $\frac{\Delta_{L}}{L}$ [$\%$]   \\
\hline
${\rm Radius}<1.5~{\rm cm}$       & $ 0.18 $ \\
$|z|<15~{\rm cm}$                 & $ 0.48 $  \\
$|cos\theta|<0.7$                 & $ 0.65 $  \\
$E_{\rm BSC}^{\rm track}>1.1~{\rm GeV}$    & $ 0.27 $  \\
$P^{\rm track}_+ ({\rm or~}P^{\rm track}_-) >0.9 E_{\rm b}$ & $1.33$ \\
Selected $z$ regions in BSC               &   $0.89$    \\ \hline
Total uncertainty in Bhabha event &   $1.83$   \\
\hline
\hline
\end{tabular}
\end{table}

\subsection{Selection of hadronic events}
In order to effectively remove the $e^+e^- \rightarrow (\gamma) e^+e^-$ and
$e^+e^- \rightarrow (\gamma) \mu^+\mu^-$ events from the selected hadronic
event sample, the hadronic events are required to have more than 2 good
charged tracks,
each of which is required to satisfy the following selection criteria:

  \begin{itemize}
        \item the charged track must be with
               a good helix fit and the number of $dE/dx$ hits per charged
               track is required to be greater than 14;

        \item the point of the closest approach to the beam line must have
              radius $\le 2.0$ cm;

        \item $|\cos \theta| \le 0.84$, where $\theta$ is the polar angle
              of the charged track;

        \item $p \le E_b+0.1\times E_b\times \sqrt {(1+E_b^2)}$, where $p$ is
              the charged track momentum and $E_b$ is the beam energy in GeV;

        \item  $2.0~ns~\le T_{TOF} \le T_p + 2.0~ns$, where $T_{TOF}$
               is the time-of-flight of the charged particle, and
               $T_p$ is the expected time-of-flight of proton with
               the given momentum;

        \item the charged track must not be identified as a muon;

        \item for the charged track, the energy deposited in BSC should be
              less than 1.0 GeV.

  \end{itemize}
In addition, the total energy deposited in BSC should be
              greater than $28\%$ of the beam energy. 
Furthermore, the selected tracks must not all point into the same hemisphere
in the $z$ direction. 
No criterion for the number of the observed photons is
applied to the selected hadronic events.

\begin{figure}
\includegraphics[width=9.5cm,height=9.0cm]
{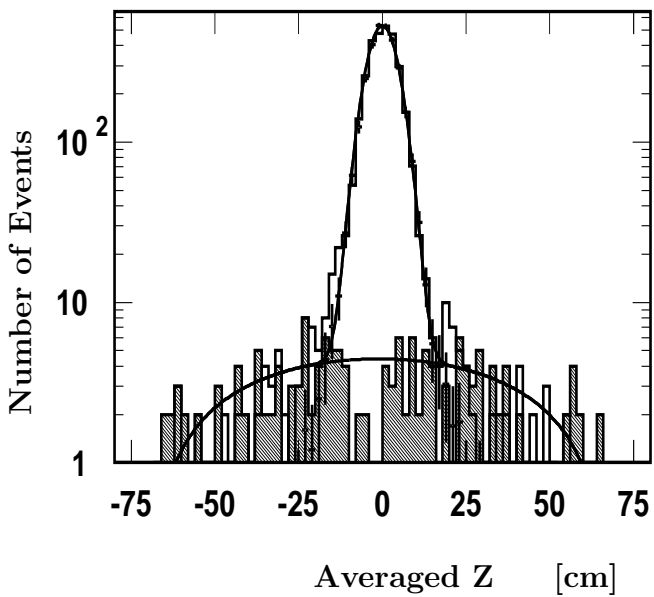}
\put(-150,10){\bf\large Averaged Z~~~~ [cm]}
\put(-265,75){\rotatebox{90}{\bf\large Number of Events}}
\caption{The distribution of the averaged $z$ of the charged tracks
which satisfy the hadronic event selection criteria,
where points with error bars show the events from the Monte Carlo sample, 
the histogram shows the events from the data,
and the shadowed histogram shows the events from the separated beam data;
the curves give the best fit to the $z$ distribution.}
\label{Zevnt_had}
\end{figure}

  Some beam-gas associated background events can also
satisfy above selection criteria.
However, the beam-gas associated background events
are produced at random $z$ positions, while the hadronic
events are produced around $z=0$. 
This characteristic can be used to distinguish the hadronic events from
the beam-gas associated background events.
To this end, the averaged $z$
of the charged tracks in each event is calculated.
Figure~\ref{Zevnt_had}
shows the distribution of the averaged $z$.  
These events are from a portion of the data taken at 3.773 GeV.
In Fig.~\ref{Zevnt_had}, the points with
error bars
show the events from the Monte Carlo sample
which is generated with the generator~\cite{zhangdh_gen}
described in Section III.C 
and simulated with the GEANT3-based Monte Carlo package~\cite{bes2_mcpck},
the histogram
shows the events from the data, and the shadowed histogram
shows the events from the separated beam data.
Using a Gaussian function plus a second
order polynomial to fit the averaged $z$ distribution of the events,
we obtain the number, $N_{\rm had}^{\rm zfit}$, of the candidates for hadronic
events.
The third column of table~\ref{lum_evt_list} lists 
$N^{\rm zfit}_{\rm had}$ obtained from the data sets 
taken at each of the energy points, where the errors 
are combined from statistical and systematic errors.
This number of candidates contains some contaminations
from some background events
such as $e^+e^- \rightarrow \tau^+\tau^-$, 
$e^+e^- \rightarrow (\gamma) e^+e^-$, 
$e^+e^- \rightarrow (\gamma) \mu^+\mu^-$
and two-photon exchange processes. The number of the background events 
can be estimated by means of Monte Carlo simulation (see Section III.E).

The systematic uncertainty in measuring the produced hadronic events
due to the hadronic event selection criteria is estimated to be about $2.5\%$. 
Table~\ref{uncertainty_had} summarizes the relative systematic uncertainties
in 
selecting 
the produced hadronic events.
\begin{table}
\centering
\vspace{0.2cm}
\caption{The relative systematic uncertainties
in measuring the produced hadronic events
due to the event selection.}
\label{uncertainty_had}
\begin{tabular}{lr}
 \hline \hline
Criterion \hspace{4.5cm}    & $\Delta_{N^{\rm prd}_{\rm had}}/N^{\rm
prd}_{\rm had}$
[$\%$]   \\ \hline
${\rm Good~helix~fit}$               & $ 0.10 $ \\
$N_{dE/dx}>14$                       &   $0.01$  \\
$V_{xy}<0.02~ \rm m$                 & $ 1.30 $ \\
$|cos\theta|<0.84$                    & $ 0.36 $  \\
$p<E_b+0.1\times E_b\times \sqrt {(1+E_b^2)}$ &  $ 0.07 $  \\
$2.0~ns~<T_{TOF}<T_p + 2.0~ns$       & $ 0.43 $     \\
$\mu {\rm Id= .false.}$              &    0.04 \\
$E_{\rm BSC}^{\rm TRK}<1.0~\rm GeV$              &  $0.56$  \\
$E_{\rm BSC}>0.28 E_{\rm b}$     &  $0.49$            \\
Same hemisphere cut                 &  $1.01$       \\
Fitting to averaged Z distribution  &  $0.65$        \\
$N_{\rm chtrk} \geq 3$       &           $1.5$ \\  \hline
Total uncertainty            &           $2.50$   \\
\hline
\hline
\end{tabular}
\end{table}

\subsection{Monte Carlo method and the efficiency $\epsilon_{had}$}

 Due to ISR (Initial State Radiation), 
the effective c.m. energy
for the $e^+e^-$ annihilation is $E_{\rm eff}=\sqrt{s(1-x)}$,
where $x\sqrt{s}/2$
is the total energy
of the emitted photons and $\sqrt{s}$ is the nominal c.m. energy.
For a certain energy point in experiment,
the experimentally observed hadronic events are not only produced
at the $\sqrt{s}$, but produced in the full energy range
from the $\sqrt{s}$ to $\sim 0.28$ GeV (for production of two pions).
To determine the efficiency for detection of hadronic events produced in the
full energy range, we developed a special Monte Carlo
generator~\cite{zhangdh_gen}
in which the initial state radiative correction to $\alpha^2$ order
is taken into account.

\begin{figure}
\includegraphics[width=9.5cm,height=9.0cm]
{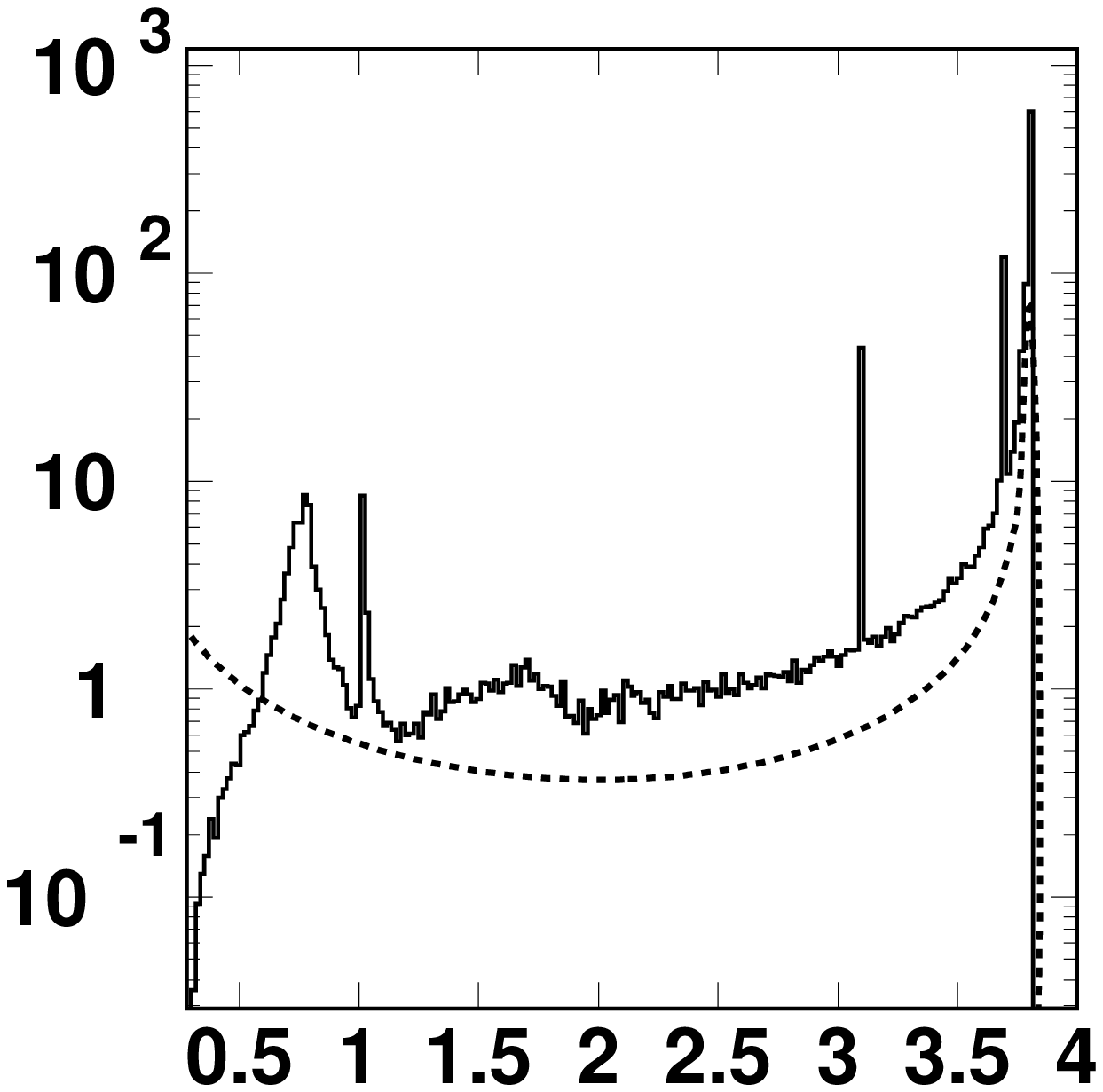}
\put(-155.0,0.0){\Large{$E_{\rm eff}$ [GeV]}}
\put(-270,85){\rotatebox{90}
     {\bf\large $d\sigma/dE_{\rm eff}$~~ [nbar/GeV]}} 
\put(-180.0,145.0){$\phi$}
\put(-195.0,155.0){$\rho$}
\put(-165.0,115.0){$\rho(1700)$}
\put(-205.0,140.0){$\omega$}
\put(-90.0,165.0){$J/\psi$}
\put(-79.0,180.0){$\psi(2S)$}
\put(-80.0,205.0){$\psi(3770)$}
\put(-135.0,70.0){$\mu^+\mu^-$}
\caption{The differential cross section for the inclusive hadronic event
production when setting
the nominal c.m. energy to be at 3.80 GeV;
the histogram shows the resonances and continuum hadronic event production;
the dashed line shows the cross sections for $\mu^+\mu^-$ pair production.
}
\label{isr_rsnc}
\end{figure}

Figure~\ref{isr_rsnc} shows the differential cross section 
$d\sigma /dE_{\rm eff}$
for the inclusive hadronic event production 
when setting
the nominal c.m. energy to be at 3.80 GeV.
At an effective c.m. energy,
the final hadronic states are produced by calling the sub-generators
such as LUND model~\cite{lund}\cite{besii_r}, 
and the resonance generators including $\psi(3770)$, $\psi(2S)$,
$J/\psi$~\cite{cjcpaper}, $\phi(1020)$, $\rho(770)$,
and $\omega(782)$ etc.
according to the corresponding lowest order cross
sections of these processes, respectively. 
The CMD-2 $\pi \pi$ production data~\cite{akhmetshin}
with Gounaris-Sakurai parameterization~\cite{gounaris}
are used to simulate the spectrum of $\rho(770)$ 
and the $\rho$-$\omega$ mixing
in the energy range below 1.2 GeV.
The resonances are set to decay into all
possible final states according to the known decay modes
and branching fractions.
These generated events are simulated with 
the GEANT3-based Monte Carlo simulation package.
The reconstructed Monte Carlo events
are then fed into the analysis program to determine the efficiencies,
$\epsilon_{had}$, for 
measurements of the observed cross
sections for inclusive hadronic event production 
at each of the three energy points.

For simulations of the inclusive hadron production,
parameters in the LUND generator are tuned using an inclusive hadronic event sample 
of $5.5\times 10^5$ events from the data taken at 3.65 GeV with the BES-II
detector. The parameters are adjusted to reproduce 
good agreeable distributions of some main kinematic variables between data
and Monte Carlo sample. The uncertainty in 
$\epsilon_{had}$ due to the adjusted parameters is estimated
to be $\sim 0.6\%$, 
while the uncertainties due to the errors of
the $\psi(3770)$ and $\psi(2S)$ resonance parameters are estimated
to be $\sim 1.5\%$ and $\sim 1.2\%$, respectively.
Combining these uncertainties in quadrature
yields the systematic uncertainty in the efficiency $\epsilon_{had}$
to be about $2\%$.

The second column
of table~\ref{effhad_obsxsct} lists the efficiencies
for detection of the inclusive hadronic events 
at three energy points 
in the case of  setting the continuum $R$ value 
to be at 2.26 (see section IV).

\begin{table}
\centering
\caption{Summary of the efficiencies for detection of the inclusive
hadronic events and the observed cross sections for 
$e^+e^- \rightarrow {hadrons}$ at three energy points, where the error in
$\epsilon_{\rm had}$ is statistical only; while the errors in
$\sigma^{\rm obs}_{\rm had}$ are statistical and point-to-point systematic,
respectively.}
\label{effhad_obsxsct}
\begin{tabular}{ccc} \hline \hline
$E_{cm}$~~~~~~ & $\epsilon_{\rm had}$ & $\sigma^{\rm obs}_{\rm had}$ \\
(GeV)~~~~~~  &   &    (nb)     \\ \hline
 3.650~~~ &  ~~~$0.4977\pm0.0022$~~~   & ~~~$18.983 \pm 0.087\pm 0.182$
\\
 3.6648~~~ &  ~~~$0.5033\pm 0.0023$~~~  & ~~~$18.299 \pm 0.199\pm 0.176$
\\
 3.773~~~ & ~~~$0.5528\pm 0.0022$~~~  & ~~~$27.680 \pm 0.056 \pm 0.266$
\\
\hline \hline
\end{tabular}
\end{table}

\subsection{Trigger efficiency}
The requirements of the trigger for recording the data on-line 
are almost the same as those used in collecting the data for 
the work~\cite{{bes2_r_prl88_y2002_p130}} and 
the work~\cite{besii_psip_rsnc}.
However, for the $\psi(3770)$ data acquisition, we slightly modified
the trigger requirements for the charged tracks, which results
in a little bit improvement in recording the two charged trak events.
The trigger efficiencies are obtained by comparing the responses to different
trigger requirements in the data taken at 3.097 GeV 
during the time period taking the data at $\sqrt{s}=3.773$ GeV.
The trigger efficiencies are measured to be 
$100.0\%$ for both the $e^+e^- \rightarrow (\gamma) e^+e^-$ and 
$e^+e^-\rightarrow {hadrons}$ events, with an uncertainty of
$^{+0.0}_{-0.5}\%$.

\subsection{Observed hadronic event cross sections}
    The observed cross section for inclusive hadronic
event production can be obtained from Eq. (3) substituting 
$N_{\rm had}(E_{{\rm cm},i})$ with $N_{\rm had}^{\rm zfit}-N_b$, 
where $N_b$ is the number of the background events,
such as $e^+e^- \rightarrow \tau^+\tau^-$, 
$e^+e^- \rightarrow (\gamma)e^+e^-$, 
$e^+e^- \rightarrow (\gamma)\mu^+\mu^-$ and
two-photon exchange processes.
The number of the background events can be estimated by
using the theoretical cross sections of these processes, 
the rates of misidentifying these processes as hadronic events
and the total integrated luminosities of the data sets, which is given by 
\begin{eqnarray}
 N_b & = & \nonumber L \times (\eta_{l^+l^-}\sigma_{l^+l^-} +
          \eta_{e^+e^-l^+l^-}\sigma_{e^+e^-l^+l^-} \\
  & &\nonumber + \eta_{e^+e^-\pi^+ \pi^-}\sigma_{e^+e^-\pi^+ \pi^-} \\
 & &      + \eta_{e^+e^-hadrons}\sigma_{e^+e^-hadrons}),
\end{eqnarray}
where $\sigma_{l^+l^-}$, $\sigma_{e^+e^-l^+l^-}$,
$\sigma_{e^+e^-\pi^+ \pi^-}$
and $\sigma_{e^+e^-hadrons}$
are the cross sections for
$e^+e^- \rightarrow l^+l^-$,
$e^+e^- \rightarrow e^+e^-l^+l^-$,
$e^+e^- \rightarrow e^+e^-\pi^+\pi^-$, and
$e^+e^- \rightarrow e^+e^-hadrons$
processes, respectively; while
$\eta_{l^+l^-}$,
$\eta_{e^+e^-l^+l^-}$,
$ \eta_{e^+e^-\pi^+ \pi^-}$, and
$\eta_{e^+e^-hadrons}$
are the corresponding misidentification rates.

In the calculation of the $\tau^+\tau^-$ cross section, 
we consider the contributions from 
$\psi(2S)$ decay,
the QED production and their interference;
we also consider the effects of the initial and final state radiative
corrections and Coulomb interaction on 
the cross section~\cite{bestau}.
For $e^+e^- \rightarrow (\gamma)e^+e^-$~\cite{kleiss_radee}, 
$e^+e^- \rightarrow (\gamma)\mu^+\mu^-$~\cite{mark_radmumu} and
two-photon processes 
$e^+e^- \rightarrow e^+e^-l^+l^-$~\cite{two_photon_berends},
the cross sections are read from respective generator outputs.
As for the estimate of the total cross section for
$e^+e^- \rightarrow e^+e^-{hadrons}$, we employ the equivalent photon
approximation formalisms to deal with the $\gamma$-$\gamma$ collision
sub-process~\cite{Brodsky}~\cite{Morgan}. 
In the sub-process, the energy dependence of the total hadronic
cross section can be described well by the formula of Donnachie-Landshoff
parameterization~\cite{Donnachie} above the three pion threshold. 
For the contribution 
in the low energy region below the three pion threshold, it is good enough to use
the simple point-like $\pi^+\pi^-$ production cross section in the
calculation~\cite{Brodsky}.

The rates of misidentifying the above processes as the hadronic events
are obtained from Monte Carlo simulation with the generators mentioned
above. For the two-photon process 
$e^+e^- \rightarrow e^+e^-hadrons$,
we use the Monte Calro generator described in ~\cite{two_photon_berends}
to simulate the process and determine
the rates of misidentifying the processes as the hadronic events.
The fourth and fifth columns of Table~\ref{lum_evt_list} give 
the estimated numbers ($n_{l^+l^-}$, $n_{e^+e^-l^+l^-}$ and $n_{e^+e^-h}$)
of the background events from 
the $e^+e^-\rightarrow l^+l^-$ and the two-photon exchange processes,
which are misidentified as
the inclusive hadronic events.

Inserting $N^{\rm zfit}_{\rm had}$, $N_b$, $L$, $\epsilon_{\rm had}$
and $\epsilon_{\rm had}^{\rm trig}$
in Eq. (3), we obtain the observed cross sections
for the inclusive hadronic event production
at each of the three energy points,
which are summarized in table~\ref{effhad_obsxsct},
where the first error is statistical and second 
point-to-point systematic error arising from the uncertainty in
$\epsilon_{e^+e^-}$ ($0.6\%$), 
uncertainty in $N_{e^+e^-}$ ($\sim (0.2\sim 0.5\%$)) 
and uncertainty in $\epsilon_{\rm had}$ ($0.5\%$). The common systematic
uncertainty is not included yet.

\section{Lowest order cross section}

\subsection{Radiative corrections}                               

To get the lowest order cross section for the inclusive hadronic event
production in $e^+e^-$ annihilation, the observed cross section has to be
corrected for the radiative effects including the initial state
radiative corrections and the vacuum polarization corrections.
The correction factor, $(1+\delta(s))$, is given by

\begin{equation}
(1+\delta(s)) = \frac{\sigma^{\rm exp}(s)}{\sigma^{\rm B}(s)},
\end{equation}
where $\sigma^{\rm exp}(s)$ is the expected cross section
and $\sigma^{\rm B}(s)$ lowest order cross section
for the inclusive hadronic event production.

The expected cross section for hadronic event production 
can be written as
\begin{equation}
  \sigma^{\rm exp}(s)= \int^{1-\frac{4m_{\pi(D)}^2}{s}}_0  dx \cdot
\frac {\sigma^{\rm B}(s(1-x))}{|1-\Pi(s(1-x))|^2} F(x,s),~~~~~~
\end{equation}
\noindent
where $\sigma^{\rm B}(s(1-x))$ is the total lowest order cross section 
in the energy range from 0.28 GeV to $\sqrt{s}$
(or from 3.729 GeV to $\sqrt{s}$ in the case of considering the $D\bar D$
production),
$F(x,s)$ is a sampling function and
$\frac{1}{|1-\Pi(s(1-x))|^2}$ is the correction factor for the
effects of vacuum polarization 
including both the leptonic and hadronic terms
in QED~\cite{Kuraev}, with
\begin{equation}
\Pi(s^{'}) = \Pi_{\rm had}(s^{'})+\Pi_{l}(s^{'}),~~~~~~
\end{equation}
the effects of hadronic vacuum polarization can be
calculated via the dispersion integral~\cite{berends} 

\begin{equation}
\Pi_{\rm had}(s^{'}) = \frac{s^{'}}{4\pi^{2}\alpha}
     \left[ {\rm PV} \int_{4m^2_{\pi}}^{\infty}
          \frac{\sigma^{\rm B}(s^{''})} { s^{'}-s^{''} }  ds^{''} -
                i\pi\sigma^{\rm B}(s^{'})\right],~~~~~~
\end{equation} 
while 
\begin{equation}
\Pi_{l}(s^{'}) = \frac{1}{2}\delta^{l^+l^-}_{vac}(s^{'}),~~~~~~                       
\end{equation} 

\begin{equation}
\delta^{l^+l^-}_{vac}(s^{'})=\frac{2\alpha}{\pi}f(\xi),~~~~~~
\end{equation}
 
\begin{equation}
f(\xi)=-\frac{5}{9}-\frac{\xi}{3}+\frac{\sqrt{1-\xi}(2+\xi)}{6}
{\rm ln}\left[\frac{1+\sqrt{1-\xi}}{1-\sqrt{1-\xi}}\right ],(\xi \le 1),~~~~~~
\end{equation}

\begin{equation}
f(\xi)=-\frac{5}{9}-\frac{\xi}{3}+\frac{\sqrt{1-\xi}
(2+\xi)}{3}{\rm tan^{-1}}\frac{1}{\sqrt{\xi-1}},(\xi >1),~~~~~~
\end{equation}
\noindent
with $\xi=\frac{4m^{2}_{l}}{s^{'}}$, where $m_l$ is the lepton mass.

In the structure function  approach  by
Kuraev and Fadin~\cite{Kuraev},
\begin{center}
\begin{equation}
F(x,s)\;=\;\beta x^{\beta-1}\delta^{V+S}+\delta^{H},~~~~~~
\end{equation}
\end{center}
where $\beta$ is the electron equivalent radiator thickness,
\begin{equation}
\beta\;=\;\frac{2\alpha}{\pi} \left(\ln \frac{s}{m^{2}_{e}}-1\right),
\end {equation}
\begin{equation}
\delta^{V+S}\;=\;1+\frac{3}{4}\beta+\frac{\alpha}{\pi}
\left(\frac{\pi^{2}}{3}-\frac{1}{2}\right)
-\frac{\beta^{2}}{24}
\left(\frac{1}{3} \ln \frac{s}{m_e^2}+ 2\pi^2 - \frac{37}{4}\right),
\end{equation}
\begin{equation}
\delta^{H}\;=\;\delta^{H}_{1}+\;\delta^{H}_{2},
\end{equation}
\begin{equation}
\delta^{H}_{1}\;=\;-\beta\left(1-\frac{x}{2}\right),
\end{equation}

\begin{equation}
\delta^{H}_{2} = \frac{1}{8}\beta^{2} \left[ 4(2-x)\ln\frac{1}{x} - 
    \frac{1+3(1-x)^{2}}{x} \ln(1-x)-6+x \right].
\end{equation}
\noindent
In above expressions, $m_e$ is the mass of electron and $\alpha$
is the fine structure constant.

For the resonances, such as 
$\psi(2S)$, $J/\psi$, $\phi$ and $\omega$, 
we use the Breit-Wigner formula
\begin{equation}
\sigma^{\rm B}(s^{'})\;=\frac{12\pi \Gamma^0_{ee} \Gamma_{h}}
   {(s^{'}-M^{2})^{2} + \Gamma^{2} M^{2}}
\end{equation}
to calculate the lowest order cross section,
where $M$ and $\Gamma$ are the mass and the total width of the resonance,
and
$\Gamma_{ee}=\Gamma^0_{ee}{|1-\Pi(s^{'})|^{-2}}$ and 
$\Gamma_{h}$ are the partial widths
to the $e^{+}e^{-}$ channel and to the inclusive hadronic final
state, respectively.
For the $\psi(3770)$ resonance, we use
\begin{equation}
\sigma^{\rm B}(s^{'})\;=\frac{12\pi \Gamma^0_{ee}
\Gamma_{tot}(s^{'})}{(s^{'}-M^{2})^{2}
+\Gamma^{2}_{tot}(s^{'}) M^{2}}
\end{equation}
to calculate the lowest order cross section,
where
$\Gamma_{\rm tot}(s^{'})$ is chosen to be energy dependent 
and normalized to the total width $\Gamma_{\rm tot}$
at the peak of the resonance.
The $\Gamma_{\rm tot}(s^{'})$ is defined as
\begin{equation}
 \Gamma_{\rm tot}(s^{'})=\Gamma_{D^0\bar D^0}(s^{'})+
          \Gamma_{D^+D^-}(s^{'})+\Gamma_{{non}-D\bar D}(s^{'}),
\end{equation}
where $\Gamma_{D^0\bar D^0}(s^{'})$, $\Gamma_{D^+D^-}(s^{'})$ and
$\Gamma_{{\rm non}-D\bar D}(s^{'})$ are the partial widths for 
$\psi(3770) \rightarrow D^0 \bar D^0$, 
$\psi(3770) \rightarrow D^+D^-$ and
$\psi(3770) \rightarrow {\rm non}-D \bar D$, respectively,
which are taken in the form
{\small
\begin{eqnarray}
\Gamma_{D^0\bar D^0}(s^{'}) & = &
          \nonumber \Gamma_0~ \theta(E_{cm}-2M_{D^0})
           \frac{(p_{D^0})^3} {(p^0_{D^0})^3}
            \frac{1+(rp_{{D^0}^0})^2} {1+(rp_{D^0})^2}B_{00}, \\
             &   & 
\end{eqnarray}
\begin{eqnarray}
\Gamma_{D^+D^-}(s^{'}) & = &
 \nonumber  \Gamma_0~ \theta(E_{cm}-2M_{D^+})
           \frac{(p_{D^+})^3} {(p^0_{D^+})^3}
            \frac{1+(rp_{D^+}^0)^2} {1+(rp_{D^+})^2}B_{+-}, \\
         &      &
\end{eqnarray}
}
and 
\begin{equation}
\Gamma_{non-D\bar D}(s^{'}) = \Gamma_0~\left[ 1
   - B_{00}
   - B_{+-}\right],
\end{equation}
where $p^0_D$ and  $p^{ }_D$ are the momenta 
of the $D$ mesons produced at the
peak of $\psi(3770)$ and 
at the c.m. energy $\sqrt{s^{'}}$, respectively;
$\Gamma_0$ is the total width of
the $\psi(3770)$ at the peak,   
and $r$ is the interaction radius of the $c\bar c$,
which is set to be 1.0 fm; 
$B_{00}$ and $B_{+-}$ are the branching fractions for 
$\psi(3770) \rightarrow D^0\bar D^0$ and
$\psi(3770) \rightarrow D^+D^-$, respectively;
$\theta(E_{cm}-2M_{D^0})$ and
$\theta(E_{cm}-2M_{D^+})$ are the step
functions to account for the thresholds of the $D\bar D$ production.

In the calculation of the lowest order cross section, the $\psi(3770)$
resonance parameters $M=3772.3\pm 1.0$ MeV, $\Gamma_0=25.5\pm3.1$ MeV
and $\Gamma_{ee}=0.224 \pm 0.031$ keV 
measured by BES Collaboration~\cite{bes_psipp_prmt}
are used.
Inserting the resonance parameters of $J/\psi$, 
$\psi(2S)$ quoted from
PDG~\cite{pdg04} and the
$R=2.26\pm 0.14$~\cite{xsct_ddbar_bes}\cite{bes2_r_prl88_y2002_p130}
for the light hadron production 
in the energy range from 2.0 to 3.0 GeV
measured by
BES Collaboration 
in Eqs. (6)--(25), we obtain the radiative correction
factors at the three energy points,
which are summarized in table~\ref{isr_0xsct_r}.
In determination of the value of $(1+\delta(s))$, the input of $R$ value
in calculating the cross section for continuum hadronic event production
affects the value of $(1+\delta(s))$. Varying the input $R$
by $\pm 10\%$ causes a variation of $\pm 1.3\%$ in $(1+\delta(s))$,
which results in a variation of the product
$\epsilon_{had} (1+\delta(s))$ by only $\pm 0.4\%$.   
The most uncontrolled cross sections in the calculation of
$(1+\delta(s))$ come from the hadronic cross sections in the energy range
from 1.2 to 2.0 GeV. However, the whole contribution 
of the cross section from this energy range is less than $5\%$ 
of the total observed cross section in our case. Since
the efficiency is quite low for detection of the hadronic events
from this energy range $(\epsilon_{had}<10\%)$, the amount of the product 
$\epsilon_{had} (1+\delta(s))$ would also be rather stable with error less
than $0.4\%$. 
Taking into account the uncertainty in 
the measured
hadronic event production and 
the errors of the resonance parameters together,
the total uncertainty in $R$ measurement due to
the calculation of $(1+\delta(s))$
is then estimated to be less than $1.5\%$ in this work.

\subsection{Lowest order cross sections and $R$ values}

The lowest order cross section for inclusive hadronic event production
is obtained by 
\begin{eqnarray}
\sigma^{\rm B}_h(s) = \frac{\sigma^{\rm obs}_{\rm h}(s)}{(1+\delta(s))},
\end{eqnarray}
where
\begin{eqnarray}
\sigma^{\rm B}_h(s) = \sigma^{\rm B}_{e^+e^-\rightarrow hadrons}(s) +
 \sum_{i}\sigma^{\rm B}_{{\rm Res},i}(s),
\end{eqnarray}
in which $\sigma^{\rm B}_{e^+e^-\rightarrow hadrons}(s)$ is 
the cross section for inclusive hadronic event production
through one photon annihilation,
$\sigma^{\rm B}_{{\rm Res},i}(s)$ is the cross section for
the i$th$ resonance, such as $J/\psi$, $\psi(2S)$, 
$\psi(3770)$ etc. which decays into hadronic final states.

To obtain the lowest order cross section 
$\sigma^0_{\rm h}(s)=R_{uds}\cdot \sigma^{\rm B}_{\mu^+\mu^-}(s)=
\sigma^{\rm B}_{e^+e^-\rightarrow hadrons}(s)$
for the hadronic event production through one photon annihilation
at the energies of 3.650 and 3.6648 GeV, 
and the lowest order cross section
$\sigma^0_{\rm h}(s)=(R_{uds}+R_{\psi(3770)})\cdot \sigma^B_{\mu^+\mu^-}=
\sigma^{\rm B}_{e^+e^-\rightarrow hadrons}(s)+
\sigma^{\rm B}_{\psi(3770)}(s)$
for both one photon annihilation and 
$\psi(3770)$ production at 3.773 GeV,
the amount of the cross section due to the resonance production at
the energies of 3.650 and 3.6648 GeV, and
the amount of the cross section due to the resonance production but
$\psi(3770)$ at 3.773 GeV have to be subtracted out.
The third column of table~\ref{isr_0xsct_r} summarizes 
the lowest order cross sections $\sigma^0_{\rm h}(s)$.
Dividing the $\sigma^0_{\rm h}(s)$ by the lowest order
cross section for $\mu^+\mu^-$ production at the same
c.m. energy, we obtain the $R$ values,
which are summarized in the fourth column of the table.
The first error in the measured lowest order cross section and the $R$
value listed in table~\ref{isr_0xsct_r} 
is statistical, the second is the point-to-point
systematic and the third is common systematic error.

The common systematic error arises from the uncertainty in
luminosity ($\sim 1.8\%$), 
in selection of hadronic event ($\sim 2.5\%$),
in Monte Carlo Modeling 
($\sim 2.0\%$),
in radiative correction ($\sim 1.5\%$)
for the measured cross sections
and $R$ values at the three energy points, 
and the uncertainty in $\psi(3770)$ resonance 
parameters ($\sim 2.7\%$)
for those at 3.773 GeV only.
Adding these uncertainties in quadrature yields the total systematic
uncertainties to be $\sim 4.0\%$ and $\sim 4.9\%$ 
for the measured hadronic cross sections and $R$ values 
for the data taken below the $D \bar D$ threshold
and at 3.773 GeV, respectively.

Averaging the $R$ values measured at the first two 
energy points (3.650 and 3.6648 GeV) 
by weighting the combined statistical and point-to-point
systematic errors, we obtain
$$ {\bar R}_{uds} = 2.224\pm 0.019\pm 0.089, $$
\noindent
where the first error is combined from statistical and 
point-to-point systematic errors, and the second is common systematic.
This $\bar R_{uds}$ 
excludes the contribution from resonances and
reflects the lowest order cross section
for the inclusive light hadronic event
production through one photon annihilation of $e^+e^-$.
So it can be directly compared with those calculated based on the pQCD.
The value is consistent with
${\bar R}_{uds}=2.26\pm0.14$
obtained by fitting those~\cite{besii_r}
measured in the energy region between 2.0 and 3.0 GeV~\cite{xsct_ddbar_bes}
and with 
$\bar R_{uds}=2.21 \pm 0.13$ obtained from fitting to the
inclusive hadronic cross sections for both the 
$\psi(2S)$ and $\psi(3770)$ resonances in the
energy region from 3.666 to 3.897 GeV~\cite{bes_psipp_prmt}.

\begin{table}
\caption{Summary of the radiative correction factors, the lowest order cross
sections and the $R$ values measured at three energy points.}
\label{isr_0xsct_r}
\begin{tiny}
\begin{tabular}{ccccc} \hline \hline
Energy    & $(1+\delta(s))$ & $\sigma^0_{\rm h}(s)$ & $R$ \\
   (GeV)  &                 &     (nb)           &     \\
\hline
3.6500   &  $1.287$ & $14.624 \pm 0.067 \pm 0.142\pm 0.590$ &
                       $2.243 \pm 0.010 \pm 0.022\pm 0.090$  \\
3.6648  &  $1.261$  & $14.151 \pm0.158  \pm 0.139\pm 0.580$ &
                       $2.188 \pm 0.024 \pm 0.022\pm 0.088$  \\
3.7730  &  $1.195$  & $23.142 \pm 0.047 \pm 0.222\pm 1.158$ &
                       $3.793\pm 0.008\pm 0.036\pm 0.190$  \\
\hline \hline
\end{tabular}
\end{tiny}
\end{table}

\begin{figure}
\includegraphics[width=9.5cm,height=8.0cm]
{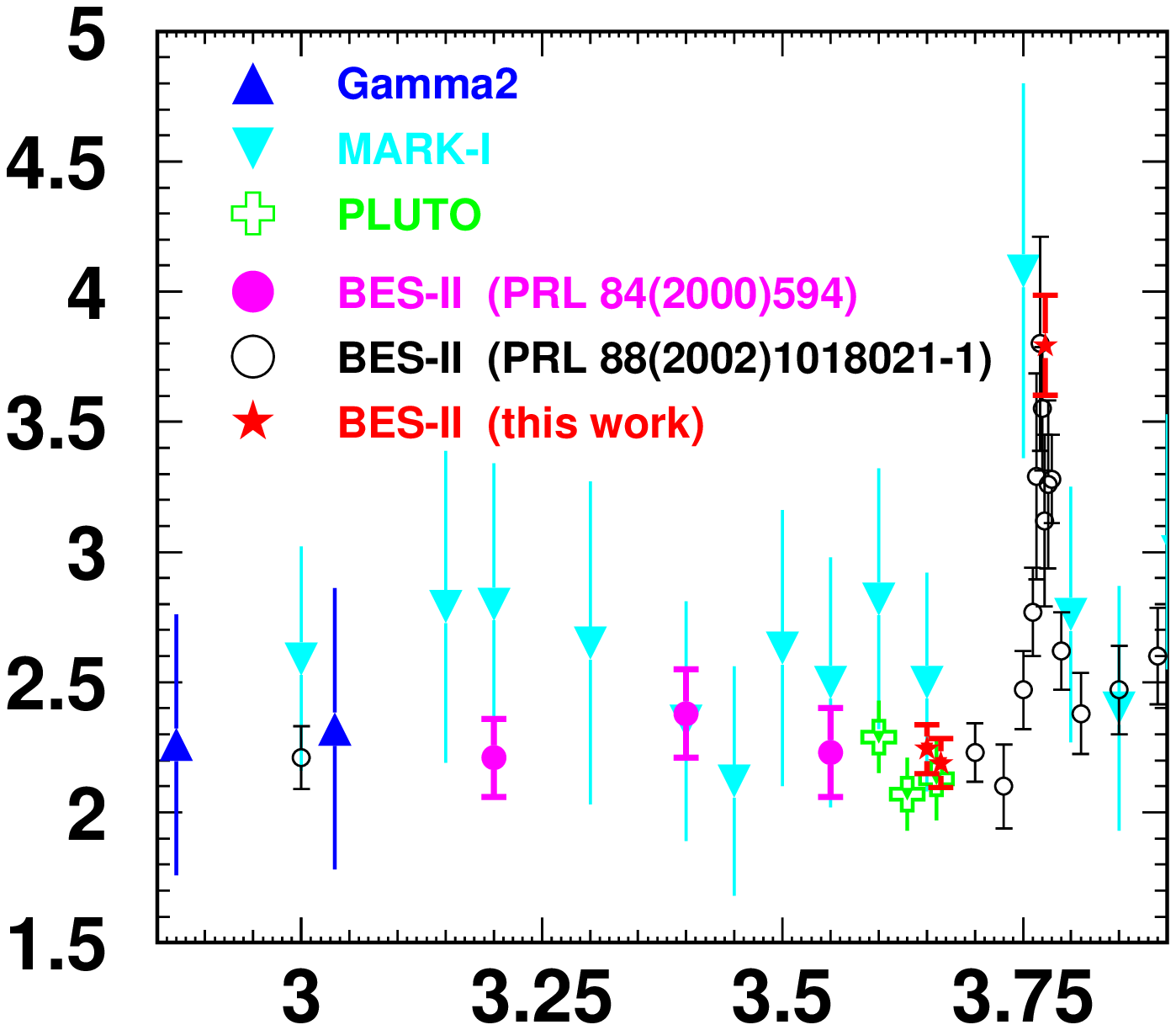}
\put(-165.0,0.0){\Large{$E_{\rm cm}$ [GeV] }}
\put(-275.0,100.0){\rotatebox{90}{\Large{$R$ Value}}}
\caption{The values of $R$ measured by BES Collaboration, MARK-I
Collaboration, $\gamma\gamma 2$ Collaboration and PLUTO Collaboration 
in the energy region between
2.85 and 3.90 GeV, where the error bars show the combined statistical
and systematic errors in quadrature.
}
\label{rplot}
\end{figure}

Figure~\ref{rplot} displays the values of $R$ from this measurement
and previous measurements by
BES Collaboration~\cite{besii_r}~\cite{bes2_r_prl88_y2002_p130},
MARK-I Collaboration~\cite{mark_i_r},
$\gamma \gamma 2$ Collaboration~\cite{gamma2}
and PLUTO Collaboration~\cite{pluto_r} in the energy region between
2.85 and 3.90 GeV. The error bars shown in the figure are obtained
by combining statistical and systematic errors in quadrature.
  
Using the measured $R$ value at 3.773 GeV listed in table~\ref{isr_0xsct_r}
and the $\bar R_{uds}$ value for light hadron production
measured 
below the $D \bar D$ threshold
we obtain the $R_{\psi(3770)}$ due to $\psi(3770)$ decays to be
\begin{equation}
 R_{\psi(3770)} = 1.569\pm 0.042 \pm 0.133,
\end{equation}
where the first error is combined from statistical and point-to-point
systematic error and the second common systematic. In
estimation of the systematic uncertainty,
we assumed that the same amount of the systematic uncertainties
in the measured values of the $R$ and the $\bar R_{uds}$ is canceled
in subtracting the $\bar R_{uds}$ from the $R$.
The corresponding lowest order cross section for $\psi(3770)$ production is
\begin{equation}
 \sigma^{\rm B}_{\psi(3770)} = (9.575\pm 0.256 \pm 0.813)~~~{\rm nb}.
\end{equation}

\section{Branching fractions for the decays $\psi(3770)\rightarrow D^0\bar D^0,
D^+D^-, D\bar D$ and for $\psi(3770)\rightarrow {\rm non}-D\bar D$}

Assuming that there are no other new structure and effects
except the $\psi(3770)$ resonance and the continuum hadron
production in the energy region
from 3.70 GeV to 3.86 GeV,
the branching fraction for $\psi(3770)\rightarrow D \bar D$ can be
determined by

\begin{equation}
BF(\psi(3770)\rightarrow D\bar D)=
   \frac{\sigma^{\rm obs}_{D\bar D}}
          {(1+\delta)_{D\bar D}~\sigma^{\rm B}_{\psi(3770)}},
\end{equation}
where $\sigma^{\rm obs}_{D\bar D}$ and
$\sigma^{\rm B}_{\psi(3770)}$ are the observed and lowest order
production cross sections for $D\bar D$ and inclusive hadronic events,
respectively; $(1+\delta)_{D\bar D}$ is the radiative correction factor 
for $D\bar D$ production. 
Inserting the $\psi(3770)$ resonance
parameters ($M=3772.3 \pm 1.0$ MeV;
$\Gamma_{\rm tot}=25.5 \pm 3.1$ MeV and
$\Gamma_{ee} = 0.224 \pm 0.031$ keV)
measured by BES Collaboration~\cite{bes_psipp_prmt} in Eqs. (6) and (7)
with combining the Eqs. (7)-(25) together, 
we obtain the radiative correction factor
\begin{equation}
 (1+\delta)_{D\bar D} = 0.764 \pm 0.014,
\end{equation}
where the error is the uncertainty arising from the errors of the
$\psi(3770)$ resonance parameters, the uncertainty
in vacuum polarization correction and the uncertainty arising from
varying the branching fraction for $\psi(3770)\rightarrow D\bar D$ from
$84\%$ to $100\%$.

BES Collaboration measured the observed cross sections for $D^0\bar D^0$
and $D^+D^-$ production at c.m. energy $\sqrt{s}=3.773$ GeV to be 
$\sigma_{D^0\bar D^0}=(3.58\pm 0.09\pm 0.31)$ nb and 
$\sigma_{D^+ D^-}=(2.56\pm 0.08\pm 0.26)$ nb~\cite{xsct_ddbar_bes}. 
These observed cross sections
were obtained by analyzing the same data set 
from which the 
$R$ value at $\sqrt{s}=3.773$ GeV is measured.

Inserting the $\sigma^{\rm B}_{\psi(3770)}$, the observed cross sections
for $D^0 \bar D^0$, $D^+D^-$, $D\bar D$ production at 3.773 GeV
and the radiative correction factor, $(1+\delta)_{D\bar D}$, in Eq. (30), 
we obtain the branching fractions for the decays
$\psi(3770) \rightarrow D^0\bar D^0,D^+D^-, D\bar D$ to be
\begin{equation}
BF(\psi(3770) \rightarrow D^0\bar D^0)=(48.9 \pm 1.2 \pm 3.8)\%,
\end{equation}
\begin{equation}
BF(\psi(3770) \rightarrow D^+ D^-)=(35.0 \pm 1.1 \pm 3.3)\%,
\end{equation}
\noindent
and
\begin{equation}
BF(\psi(3770) \rightarrow D\bar D)=(83.9 \pm 1.6 \pm 5.7)\%,
\end{equation}
which results in the non-$D\bar D$ branching fraction of $\psi(3770)$ to be
\begin{equation}
BF(\psi(3770) \rightarrow {\rm non}-D\bar D)=(16.1 \pm 1.6 \pm 5.7)\%,
\end{equation}
where the first error is statistical and the second systematic arising
from uncanceled systematic uncertainties. The uncanceled relative
systematic uncertainties are 
$\sim 6.2\%$, $\sim 8.4\%$ and $\sim 3.2\%$ for the 
$\sigma^{\rm obs}_{D^0\bar D^0}$, $\sigma^{\rm obs}_{D^+D^-}$
and the $\sigma^{\rm B}_{\psi(3770)}$, respectively.
The systematic error also includes the 
common uncertainty of $\sim 2.7\%$
arising from the statistical uncertainty 
in the measured lowest order cross section for $\psi(3770)$ production and
the uncertainty ($\sim 1.8\%$) in radiative correction factor
$(1+\delta)_{D\bar D}$.
Table~\ref{uncancel_syserr} summarizes sources of the uncanceled systematic
uncertainties for the measured $\sigma^{\rm obs}_{D\bar D}$
and $\sigma_{\psi(3770)}$.
The uncertainties in luminosity ($\sim 1.8\%$), in $\psi(3770)$ resonance
parameters ($\sim 2.7\%$) and in radiative correction ($\sim 1.5\%$) are
canceled out in the estimation of the systematic uncertainty in the
measured branching fractions.
\begin{table}
\caption{Sources of the uncanceled systematic uncertainties in the measured
cross sections for $D\bar D$ and $\psi(3770)$ production, 
where upper $^N$ and upper $^C$ mean the uncertainties for
the neutral mode $D^0\bar D^0$ and 
the charged mode $D^+D^-$, respectively; 
for the uncertainties in the measured cross 
section for $D\bar D$ production, 
please refer to the reference \cite{xsct_ddbar_bes};
we here re-estimate the uncertainty in the combined tracking and kinematic
fit in selection of $D$ events to be $\sim 4.0\%$.}
\label{uncancel_syserr}
\begin{center}
\begin{tabular}{cccc} \hline \hline
 Source &$\frac{\Delta \sigma^{\rm obs}_{D\bar D}}
                                   {\sigma^{\rm obs}_{D\bar D}}$ &
 Source &$\frac{\Delta \sigma_{\psi(3770)}}{\sigma_{\psi(3770)}}$ \\
        &    [$\%$]   &  &  [$\%$]  \\
\hline
Particle ID & $\sim 1.5$   & Monte Carlo Modeling  & $\sim 2.0$  \\
Tracking $\&$ K.F.
        & $\sim 4.0$  & Hadron selection & $\sim 2.5$   \\
F. P.       & $\sim 3.0$  &        &     \\
MC Statistics &$\sim 0.6$  &     &     \\
Br for $D^0$    & $\sim 3.2$   &        &    \\
(Br for $D^+$)  & ($\sim 6.5$) &        &   \\
\hline
Total  &  $\sim 6.2$~$^{N}$ & Total    & $\sim 3.2$  \\
       & ($\sim 8.4$)~$^C$ &     &             \\
\hline \hline
\end{tabular}
\end{center}
\end{table}

\section{Discussion about interference effects}
The measured $R$ values discussed in above sections are obtained based on
the same treatment
on the measurements of inclusive hadronic cross sections 
in which no interference between the inclusive hadronic
final states of the resonance decays and the inclusive hadronic final states 
from non-resonance annihilation of $e^+e^-$ 
is taken into
account~\cite{besii_r}\cite{bes2_r_prl88_y2002_p130}\cite{r_and_prmts_no_intf}.
However, 
since the c.m. energies of 3.650 GeV, 3.6648 GeV and 3.773 GeV are close to
the $\psi(2S)$ resonance, there may be interference effects 
between the final hadronic states from the 
$\psi(2S)$ electromagnetic decays
and the continuum hadron production in annihilation of $e^+e^-$.
These interference effects distort the line shape of the
continuum hadron production cross section around the  
$\psi(2S)$ peak.
With the definition of the $R$ given in Eq. (2), we can estimate the
destructive/constructive amount
of the cross section due to the interference effects,
which is given by
\begin{equation} 
\sigma^{\rm interf}_{\rm had} = R~\sigma^{\rm interf}_{\mu^+\mu^-}.
\end{equation}
The destructive/constructive amounts of the cross sections given in $R$
are estimated to be 
$-0.0581$, $-0.1026$ and $+0.0219$
at 3.650 GeV, 3.6648 GeV and 3.773 GeV, respectively.
After correcting the cross section $\sigma^{\rm 0}_{h}(s)$ for
the destructive/constructive amounts due to the interference effects,
we obtained the lowest order cross section $\sigma^{0~{\rm crr}}_{\rm h}(s)$.  
The third column of table~\ref{isr_0xsct_r_crrct_intf} summarizes the 
$\sigma^{0~{\rm crr}}_{\rm h}(s)$. 
In the case of considering the interference effects, 
the correction factor $(1+\delta(s))$ is also changed. 
The second column of table~\ref{isr_0xsct_r_crrct_intf} lists
the correction factor at the c.m. energies.
Dividing the $\sigma^{0~{\rm crr}}_{\rm h}(s)$ by the lowest order
cross section for $\mu^+\mu^-$ production at the same
c.m. energy, we obtain the $R$ values,
which are summarized in the fourth column of the table.
The errors are statistical, the point-to-point
systematic and the common systematic as discussed before.
\begin{table}
\caption{Summary of the radiative correction factors, the lowest order cross
sections and the $R$ values measured at three energy points, where
the interference effects between the 
$\psi(2S)$ electromagnetic decays
and hadron production through
non-resonant annihilation of $e^+e^-$
are taken into account.}
\label{isr_0xsct_r_crrct_intf}
\begin{tiny}
\begin{tabular}{ccccc} \hline \hline
Energy    & $(1+\delta(s))$ & $\sigma^{0~\rm crr}_{\rm h}(s)$ & $R$ \\
   (GeV)  &                 &     (nb)           &     \\   
\hline
3.6500   & $1.300$ & $14.855 \pm 0.067 \pm 0.140\pm 0.584$ &
                       $2.279 \pm 0.010 \pm 0.022\pm 0.089$  \\
3.6648  &  $1.285$  & $14.543 \pm0.155  \pm 0.137\pm 0.570$ &  
                       $2.249 \pm 0.024 \pm 0.021\pm 0.086$  \\
3.7730  &  $1.198$  & $22.951 \pm 0.047 \pm 0.222\pm 1.155$ &  
                       $3.762\pm 0.008\pm 0.036\pm 0.189$  \\  
\hline \hline
\end{tabular}
\end{tiny}   
\end{table}  

The weighted average of the $R$ values measured at the first two
energy points is
$$ {\bar R}_{uds} = 2.268\pm 0.019\pm 0.091, $$
\noindent
where the first error is combined from statistical and point-to-point
systematic errors, and the second is common systematic.

  Following the same procedure as discussed in Sections IV and Section V,
we obtained the lowest order cross section for $\psi(3770)$ production
to be
\begin{equation}
 \sigma^{\rm B}_{\psi(3770)} = (9.113 \pm 0.255 \pm 0.805)~~~{\rm nb},
\end{equation}
and the branching fractions for $\psi(3770)$ decays to be
\begin{equation}
BF(\psi(3770) \rightarrow D^0\bar D^0)=(51.4 \pm 1.3 \pm 4.0)\%,
\end{equation}
\begin{equation}
BF(\psi(3770) \rightarrow D^+ D^-)=(36.8 \pm 1.1 \pm 3.5)\%,
\end{equation}
\begin{equation}
BF(\psi(3770) \rightarrow D\bar D)=(88.2 \pm 1.7 \pm 6.0)\%,
\end{equation}
\noindent 
and
\begin{equation}
BF(\psi(3770) \rightarrow {\rm non}-D\bar D)=(11.8 \pm 1.7 \pm 6.0)\%,
\end{equation}
where the errors are statistical and the systematic arising
from some uncanceled systematic uncertainties. 

\section{Summary}

From the analysis of the data taken at 3.650, 3.6648 and 3.773 GeV in
$e^+e^-$ annihilation, we
measured the $R$ values at the energy points, 
which are $R=3.793 \pm 0.037 \pm 0.190$ at 3.773 GeV and 
$\bar R_{uds}=2.224\pm 0.019\pm 0.089$ at $\sqrt{s}$
below the $D\bar D$ threshold.
Based on the measured $R$ values, 
we determined the lowest order cross section for $\psi(3770)$ production 
at 3.773 GeV to be 
$\sigma^{\rm B}_{\psi(3770)} = (9.575\pm 0.256 \pm 0.813)~{\rm nb}$,
the branching fractions
for $\psi(3770)$ decays to be
$BF(\psi(3770) \rightarrow D^0\bar D^0)=(48.9 \pm 1.2 \pm 3.8)\%$,
$BF(\psi(3770) \rightarrow D^+ D^-)=(35.0 \pm 1.1 \pm 3.3)\%$ and
$BF(\psi(3770) \rightarrow D\bar D)=(83.9 \pm 1.6 \pm 5.7)\%$,
which result in the non-$D\bar D$ branching fraction of $\psi(3770)$ to be
$BF(\psi(3770) \rightarrow {\rm non}-D\bar D)=(16.1 \pm 1.6 \pm 5.7)\%$.
These branching fractions for $\psi(3770)$ decays are measured for the first
time. 

\vspace{3mm}

\begin{center}
{\small {\bf ACKNOWLEDGEMENTS}}
\end{center}

\vspace{0.4cm}

   The BES collaboration thanks the staff of BEPC for their hard efforts.
This work is supported in part by the National Natural Science Foundation
of China under contracts
Nos. 19991480,10225524,10225525, the Chinese Academy
of Sciences under contract No. KJ 95T-03, the 100 Talents Program of CAS
under Contract Nos. U-11, U-24, U-25, and the Knowledge Innovation Project
of CAS under Contract Nos. U-602, U-34(IHEP); by the
National Natural Science
Foundation of China under Contract No.10175060(USTC),and
No.10225522(Tsinghua University).

\end{document}